\pdfoutput=1
\documentclass[conference,9pt]{IEEEtran}
\IEEEoverridecommandlockouts
\usepackage{cite}
\usepackage{amsmath,amssymb,amsfonts}
\usepackage{graphicx}
\usepackage{textcomp}
\usepackage{xcolor}
\usepackage[T1]{fontenc}
\usepackage{booktabs}
\usepackage{array}
\usepackage{url}
\usepackage{microtype}
\usepackage[hidelinks]{hyperref}

\newcommand{\sass}[1]{\texttt{\small #1}}
\newcommand{\EX}{\textup{\textsc{exempt}}}
\newcommand{\TILE}{\textup{\textsc{tile}}}
\newcommand{\START}{\textup{\textsc{start}}}
\newcommand{\END}{\textup{\textsc{end}}}

\hypersetup{
  pdftitle={Dissecting an Unexplored SASS Control-Code Field: Barrier Exemption and Programmable Warp Priority},
  pdfauthor={Ziwen Wang, Qi Luo, Kunlin Li, Changxi Liu, Dongsheng Wang, Yun Chen}
}

\begin{document}

\title{Dissecting an Unexplored SASS Control-Code Field:\\ Barrier Exemption and Programmable Warp Priority}

\author{\IEEEauthorblockN{Ziwen Wang\textsuperscript{1},
Qi Luo\textsuperscript{1},
Kunlin Li\textsuperscript{1},
Changxi Liu\textsuperscript{1,*},
Dongsheng Wang\textsuperscript{2},
Yun Chen\textsuperscript{1,*}}
\IEEEauthorblockA{\textsuperscript{1}\textit{The Hong Kong University of Science and Technology (Guangzhou)} \\
\textsuperscript{2}\textit{Tsinghua University} \\
\{zwang578, qluo615, kli082\}@connect.hkust-gz.edu.cn,
\{changxiliu, yunchen\}@hkust-gz.edu.cn,
wds@tsinghua.edu.cn \\
\textsuperscript{*}Corresponding authors}}
\maketitle

\begin{abstract}
Every instruction in NVIDIA's native GPU ISA (SASS) carries a control code that guides hardware scheduling, and prior work has decoded its known fields to optimize GPU kernels.
However, one control-code field remains unexplored: it appears over 500000 times in NVIDIA's closed-source libraries such as cuBLAS and cuDNN, yet \sass{nvcc} never emits it when compiling user programs.
This paper presents the first systematic study of this control-code field across GPU architectures from Volta to Blackwell.
Using instruction-level microbenchmarks, we show that the field controls two independent hardware mechanisms: barrier exemption (\EX{}), which allows instructions blocked by a pending barrier to issue early during the barrier wait, and programmable warp priority (\START{}, \TILE{}, and \END{}), which moves warps across three priority levels in the hardware scheduler.
We identify the conditions under which each mechanism improves GPU kernel performance and implement a post-compilation binary optimizer that places these marks directly in compiled binaries without modifying source code or the compiler.
Evaluated across six GPUs spanning four architectures, our optimizations speed up representative kernels by up to 31.9\%, and the optimizer accelerates production kernels in \sass{llama.cpp} and \sass{vLLM} by up to 5.4\% and end-to-end LLM inference by up to 3.6\%.
\end{abstract}

\section{Introduction}
\label{sec:intro}

Programs running on NVIDIA GPUs are ultimately compiled into SASS, the native machine instruction set executed by the hardware.
Since the Volta architecture, every 128-bit SASS instruction encodes not only an opcode and operands, but also a 23-bit control code that directly guides how the hardware warp schedulers issue and synchronize instructions.
Because NVIDIA does not document SASS control codes, prior work has reverse-engineered their fields, including the stall count, yield bit, scoreboards, and operand-reuse flags~\cite{jia18volta,huerta25cores}, and tuned these fields to optimize GPU kernel performance~\cite{zhang17baremetal,he25cuasmrl}.

However, one 3-bit field in this control code remains unexplored.
Early reverse-engineering studies treated these bits solely as operand-reuse flags~\cite{jia18volta}, and while F2Asm~\cite{gai26f2asm} and DocumentSASS~\cite{documentsass} recently noticed the field's existence, neither studied its hardware behavior.
To determine whether this field is merely an unused encoding artifact, we inspect NVIDIA's closed-source libraries, such as \sass{cuBLAS} and \sass{cuDNN}, and find that it appears over 500000 times.
By contrast, neither CUDA nor PTX exposes this field, and \sass{nvcc} never emits it when compiling user programs, leaving its performance benefits restricted to NVIDIA's closed-source libraries.
This paper presents the first systematic study of this control-code field.

Using instruction-level microbenchmarks, we show that this field controls two independent hardware mechanisms.
The first mechanism is \emph{barrier exemption} (\EX{}), which allows instructions blocked by a pending barrier to issue early while waiting for other warps to reach the barrier, providing a hardware realization of the fuzzy barrier~\cite{gupta89fuzzy}.
The second mechanism is \emph{programmable warp priority} (\START{}, \TILE{}, and \END{}), which moves warps across three priority levels (high, normal, and low) in the warp scheduler, allowing software to boost or demote warp priority at the instruction level.

Both mechanisms can be used to optimize GPU kernel performance: \EX{} reduces barrier synchronization overhead by letting early warps execute independent instructions during barrier waits, while \TILE{} and \END{} balance progress across multiple CTAs on an SM to keep enough warps ready to hide memory latency, speeding up representative kernels by up to 31.9\% and 21\%, respectively.
To apply both optimizations to existing CUDA programs, we implement a post-compilation binary optimizer that places \EX{}, \TILE{}, and \END{} directly into the control codes of compiled binaries without modifying source code or the compiler.
Evaluated across six GPUs spanning four architectures, the optimizer accelerates production kernels in \sass{llama.cpp} and \sass{vLLM} by up to 5.4\% and end-to-end LLM inference by up to 3.6\%.

In summary, this paper makes the following contributions:
\begin{itemize}
\item We present the first systematic study of an unexplored control-code field in NVIDIA GPUs from Volta to Blackwell, uncovering a hardware fuzzy barrier (\EX{}) and three-level programmable warp priority (\START{}, \TILE{}, and \END{}).
\item We identify the conditions under which each mechanism improves GPU kernel performance, speeding up representative kernels by up to 31.9\% and 21\%, respectively.
\item We implement a post-compilation binary optimizer that places these marks directly in compiled binaries without modifying source code or the compiler, and evaluate it across six GPUs spanning four architectures on \sass{llama.cpp} and \sass{vLLM}, accelerating production kernels by up to 5.4\% and end-to-end LLM inference by up to 3.6\%.
\end{itemize}

\section{Background and Motivation}
\label{sec:bg}

\subsection{SASS Control Codes and Deferred Barriers}
\label{sec:control}

\begin{figure}[t]
\centering
\includegraphics[width=\columnwidth]{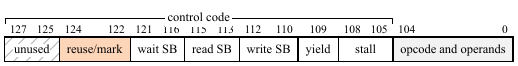}
\caption{Bit layout of a 128-bit SASS instruction. Bits \texttt{124:122} hold either operand-reuse flags or
one of the marks \EX{}, \START{}, \TILE{}, and \END{}. SB: scoreboard.}
\label{fig:control}
\end{figure}

NVIDIA GPUs execute SASS, their native instruction set: \sass{nvcc} compiles CUDA source code to the virtual ISA PTX, and the closed-source assembler \sass{ptxas} translates PTX into SASS and packages it into a binary (\sass{cubin}).
Since Volta, every 128-bit SASS instruction consists of a 105-bit opcode and operand word (bits \texttt{104:0}) and a 23-bit control code (bits \texttt{127:105}) set by \sass{ptxas} to guide hardware scheduling (Fig.~\ref{fig:control})~\cite{jia18volta,huerta25cores}.
Prior work has decoded four scheduling fields in this control code.
The \emph{stall count} (bits \texttt{108:105}) specifies how many cycles (0 to 15) the scheduler must wait after issuing the current instruction before issuing the next instruction from the same warp.
The \emph{yield bit} (bit \texttt{109}) indicates whether the scheduler should keep issuing from the current warp on the next cycle or may switch to another ready warp.
\emph{Scoreboards} (bits \texttt{121:110}, comprising a write scoreboard in bits \texttt{112:110}, a read scoreboard in bits \texttt{115:113}, and a 6-bit wait scoreboard mask in bits \texttt{121:116}) track variable-latency operations such as memory accesses, where a producer instruction sets a scoreboard and a dependent instruction (or an explicit \sass{DEPBAR} instruction) waits until that scoreboard clears.
Finally, \emph{operand-reuse flags} (bits \texttt{124:122}) hint the hardware to cache up to three source register values in the register cache for reuse by subsequent instructions.

When a kernel launches on a GPU, its threads are grouped into CTAs (thread blocks) distributed across the GPU's streaming multiprocessors (SMs), with each CTA partitioned into 32-thread warps.
Each SM contains four warp schedulers, and on every cycle each scheduler selects one ready warp to issue a SASS instruction.
Because the warps of a CTA execute independently, coordinating them (e.g., when exchanging data through shared memory) requires a barrier such as CUDA's \sass{\_\_syncthreads()}, which since Volta compiles to the SASS deferred barrier instruction \sass{BAR.SYNC.DEFER\_BLOCKING}.
Unlike an older blocking \sass{BAR.SYNC}, a deferred barrier does not stop an arriving warp at the \sass{BAR} instruction itself while the barrier is \emph{pending} (i.e., while other warps in the CTA have not yet reached the barrier).
Instead, the warp continues issuing subsequent instructions after the \sass{BAR} and stalls only when it reaches an instruction that the pending barrier blocks (\S\ref{sec:exempt}).

\subsection{An Unexplored Control-Code Field}
\label{sec:corpus}

Prior reverse-engineering studies~\cite{jia18volta} treated bits \texttt{124:122} of the control code solely as operand-reuse flags.
In fact, across every GPU architecture from Volta to Blackwell, bits \texttt{124:122} encode either operand-reuse flags or a scheduling mark depending on the instruction type and the yield bit (Fig.~\ref{fig:control}): only arithmetic instructions support operand reuse, using bits \texttt{124:122} as reuse flags when the yield bit is unset and as a mark when the yield bit is set, whereas all other instructions (such as memory accesses and branches) do not use operand reuse and always encode a mark in bits \texttt{124:122}.
This field takes four marks (with \texttt{000} indicating no mark): \EX{} (\texttt{101}), \START{} (\texttt{001}), \TILE{} (\texttt{010}), and \END{} (\texttt{100}).
Recent work has noticed this field without studying its hardware behavior: F2Asm~\cite{gai26f2asm} observed two \sass{DEPBAR} instructions with identical disassembly text that differed only in bits \texttt{122} and \texttt{124} and preserved those bits as opaque state, while DocumentSASS~\cite{documentsass} listed the field's names and speculated only that it might relate to variable-latency instructions.

To determine whether this field is merely an unused encoding artifact or is actively used in production binaries, we inspect 25806 \sass{cubin}s across NVIDIA's closed-source libraries (\sass{cuBLASLt}, \sass{cuDNN}, \sass{cuSOLVER}, \sass{cuSPARSELt}, \sass{NCCL}, \sass{NPP}, and \sass{cuFFT}) as well as user programs compiled by \sass{nvcc} (PyTorch, FlashAttention, vLLM, and \sass{llama.cpp}).
In NVIDIA's libraries, the field is used extensively: \EX{} appears 503175 times (including across the GEMM kernels of \sass{cuBLAS}), and \START{} and \END{} appear as 2085 pairs in the convolution kernels of \sass{cuDNN}, whereas \TILE{} never appears.
By contrast, code compiled by \sass{nvcc} never emits any of these marks.
This contrast shows that the field controls active hardware mechanisms that NVIDIA's libraries use for performance optimization, yet because its behavior is undocumented and \sass{nvcc} never emits it, user programs compiled with \sass{nvcc} cannot benefit from it, motivating us to uncover its hardware mechanisms and apply them automatically to existing programs.

\section{Barrier Exemption (\EX)}
\label{sec:exempt}

\subsection{Hardware Mechanism}
\label{sec:exempt-behavior}

\begin{figure}[t]
\centering
\includegraphics[width=\columnwidth]{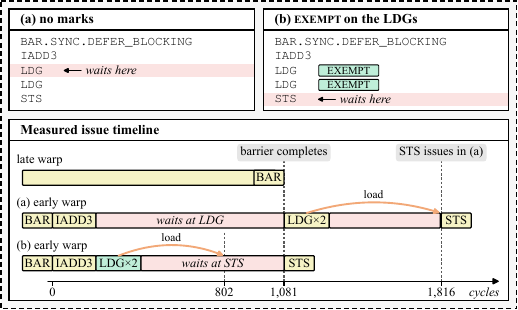}
\caption{The \EX{} microbenchmark and its issue timeline measured on an H20: (a)~without marks,
(b)~with \EX{} on the \sass{LDG}s. Boxes mark the cycle at which an instruction issues; their widths
are not to scale.}
\label{fig:exempt}
\end{figure}

To determine the effect of \EX{} while a deferred barrier is pending, we construct a SASS microbenchmark with one CTA of two warps (Fig.~\ref{fig:exempt}).
An \emph{early warp} reaches \sass{BAR.SYNC.DEFER\_BLOCKING} first, while a \emph{late warp} arrives 1081 cycles later, keeping the barrier pending during that window.
After the \sass{BAR}, the early warp executes \sass{IADD3} to compute an address, two global loads (\sass{LDG}) to fetch data from global memory, and \sass{STS} to write the data to shared memory.
We compare running this sequence without marks (Fig.~\ref{fig:exempt}(a)) and with \EX{} on the two \sass{LDG}s (Fig.~\ref{fig:exempt}(b)), recording the issue cycle of each instruction using the hardware clock.

Without marks (Fig.~\ref{fig:exempt}(a)), the early warp issues \sass{IADD3} right after the \sass{BAR} at cycle~35, but stalls at the first \sass{LDG}.
The two \sass{LDG}s wait until cycles 1083 and 1087, just after the late warp completes the barrier at cycle 1081.
Because \sass{STS} depends on the loaded data, it must then wait for the loads to return from global memory and does not issue until cycle 1816.
With \EX{} on both \sass{LDG}s (Fig.~\ref{fig:exempt}(b)), the early warp does not wait at the loads.
It issues the two marked \sass{LDG}s at cycles~39 and~43, before the late warp reaches the barrier, and stalls only when it reaches \sass{STS}.
The loads return from global memory at cycle~802, before the barrier completes at cycle 1081.
\sass{STS} then issues at cycle 1083, 733~cycles earlier than without marks.
When we mark only the first \sass{LDG}, that load issues at cycle~39, and the warp stalls at the second, unmarked \sass{LDG} until cycle 1083.

These timelines show how the hardware handles a pending deferred barrier and what \EX{} changes.
While the barrier is pending, a warp continues issuing instructions that the barrier does not block (such as \sass{IADD3}) and stalls when it reaches the first instruction that the barrier blocks (such as the first \sass{LDG} in Fig.~\ref{fig:exempt}(a)).
Placing \EX{} on a blocked instruction allows it to issue without waiting for the pending barrier, so the warp continues issuing subsequent instructions until it reaches the first unmarked instruction that the barrier blocks (such as \sass{STS} in Fig.~\ref{fig:exempt}(b)).
This mechanism is a hardware realization of the fuzzy barrier~\cite{gupta89fuzzy}, which separates barrier arrival from the completion wait so that an arriving warp can execute compiler-selected instructions while waiting for other warps to reach the barrier.
Thus, current NVIDIA GPUs already support fuzzy barriers in hardware and use the \EX{} mark in the control code to select which blocked instructions can issue during a pending barrier, yet user programs cannot benefit from this mechanism because \sass{nvcc} never emits \EX{}.

\begin{table}[t]
\caption{Barrier blocking and \EX{} release by execution pipeline on Hopper.}
\label{tab:block}
\centering\footnotesize
\setlength{\tabcolsep}{4.5pt}
\newcommand{\op}[1]{\texttt{#1}}
\begin{tabular}{@{}llccr@{}}
\toprule
Pipeline & Examples & Blocked & Released & Marks \\
\midrule
ALU                  & \op{IADD3 LOP3}      & --         & -- & 0 \\
FMA                  & \op{FFMA IMAD}       & --         & -- & 0 \\
FP64                 & \op{DFMA DADD}       & --         & -- & 0 \\
Uniform              & \op{UIADD3 ULOP3}    & --         & -- & 0 \\
Tensor (MMA)         & \op{HMMA DMMA}       & --         & -- & 0 \\
\midrule
Tensor (wgmma)       & \op{HGMMA IGMMA}     & \checkmark & \checkmark & 0 \\
SFU                  & \op{MUFU F2F}        & \checkmark & \checkmark & 23705 \\
LSU                  & \op{LDG STS}         & \checkmark & \checkmark & 302145 \\
TEX                  & \op{TEX SULD}        & \checkmark & \checkmark & 0 \\
ADU                  & \op{BAR LDC}         & \checkmark & \checkmark & 10776 \\
CBU                  & \op{BSSY EXIT}       & \checkmark & \checkmark & 1505 \\
TMA                  & \op{UTMALDG UBLKCP}  & \checkmark & \checkmark & 0 \\
Other                & \op{NOP DEPBAR}      & \checkmark & \checkmark$^{*}$ & 165044 \\
Fence                & \op{MEMBAR}          & \checkmark & $\times$ & 0 \\
\bottomrule
\end{tabular}\\[2pt]
\parbox{\columnwidth}{\footnotesize $^{*}$\,With the yield bit.
\op{ACQBULK}, \op{ERRBAR}, and \op{WARPGROUP} (Other) are not blocked.
Marks: \EX{} marks in NVIDIA's libraries.}
\end{table}

To determine which instructions a pending barrier blocks and which \EX{} releases across the instruction set, we test each SASS instruction after the pending barrier with and without \EX{} across four GPU architectures (Ampere, Ada, Hopper, and Blackwell), and summarize the Hopper results by execution pipeline in Table~\ref{tab:block}.
Instructions executed on the fixed-latency arithmetic pipelines (ALU, FMA, FP64, Uniform, and Tensor MMA) are never blocked by a pending barrier.
Instructions executed on the remaining pipelines (Tensor \sass{wgmma}, SFU, LSU, TEX, ADU, CBU, and TMA) are blocked by a pending barrier and released by \EX{}.
A small set of other blocked instructions (\texttt{Other} in Table~\ref{tab:block}, such as \sass{NOP}, \sass{DEPBAR}, unconditional \sass{BRA}, and \sass{ULDC}) require setting the yield bit alongside \EX{} to be released.
Finally, memory fences (\sass{MEMBAR}) are blocked by a pending barrier and cannot be released even with \EX{}.
This mechanism holds across the other architectures, except that on architectures where FP64 instructions have variable latency (Ampere, Ada, and Blackwell), FP64 instructions are also blocked by a pending barrier and released by \EX{}.
In all cases, \EX{} only exempts an instruction from waiting for a pending deferred barrier, whereas the warp still obeys normal instruction dependencies such as stall counts and scoreboard waits for prior memory accesses.
These rules explain the distribution of \EX{} marks across NVIDIA's libraries in the last column of Table~\ref{tab:block} (\S\ref{sec:corpus}).
Instructions executed on the arithmetic pipelines carry zero marks because a pending barrier never blocks them.
By contrast, in kernels such as \sass{cuBLAS} GEMM, NVIDIA marks blocked instructions in LSU (\sass{LDG}), Other (\sass{BRA}), SFU (\sass{F2F}), and ADU (\sass{LDC}) between a deferred barrier and the first shared-memory access to overlap fetching and converting the next tile with the barrier wait.

\subsection{Performance Optimization}
\label{sec:exempt-opt}

Because \EX{} allows instructions blocked by a pending deferred barrier to issue early, marking blocked instructions after a barrier with \EX{} reduces barrier stalls by letting early warps continue executing subsequent instructions while waiting for slower warps.
Among the blocked categories in Table~\ref{tab:block}, marking SFU operations (\sass{MUFU}, \sass{F2F}), control-flow instructions in CBU and Other (\sass{BSSY}, \sass{BRA}), and constant or special-register reads in ADU (\sass{LDC}, \sass{S2R}) lets an early warp issue and execute both those instructions and any arithmetic instructions trapped behind them during the barrier wait.
Because these instructions operate only on the warp's own registers or read-only constants, which no other warp can modify, executing them early never changes their results.
For memory accesses in LSU, marking thread-private local-memory accesses (\sass{LDL}, \sass{STL}) and accesses to global or shared addresses that no other warp writes across the barrier (such as read-only \sass{LDG}s) likewise does not change their results, and lets an early warp issue them early to overlap their memory latency with the barrier wait (as in Fig.~\ref{fig:exempt}(b)).
Conversely, if another warp reads or writes the same shared or global address across the barrier, issuing the access early would read or overwrite data before that warp reaches the barrier.
To preserve correctness, such accesses must remain unmarked so that the early warp stops there and waits for the barrier (as at \sass{STS} in Fig.~\ref{fig:exempt}(b)).

However, reducing barrier stalls for a waiting warp does not always shorten total kernel execution time.
If enough other warps on the SM are ready to fill the issue slots during the barrier wait, \EX{} reduces barrier stalls without making the kernel faster.
Marking instructions with \EX{} therefore speeds up a kernel only when the SM schedulers lack other ready warps to issue during the barrier wait, which happens in two cases.

First, warps within a CTA may have uneven workloads~\cite{liu16baws}.
When most warps finish early and stall at the barrier waiting for a few slow warps, only those few slow warps remain active, leaving the SM schedulers without enough ready warps to keep the issue slots busy.
We test this case on the H20 with an EmbeddingBag~\cite{paszke19pytorch} kernel: in each iteration, each warp loads and sums a bag of rows from a read-only embedding table in global memory, writes its partial sum to shared memory, and synchronizes at a barrier.
Marking the instructions after the barrier up to the next iteration's shared-memory store lets early warps start loading and summing their next bag while waiting at the barrier.
When every bag has 32 rows (balanced workload), all warps reach the barrier together and marking has no effect.
By contrast, when 90\% of bags have 16 rows and 10\% have 176 rows (imbalanced workload), warps with 16-row bags execute their next bag while waiting for warps with 176-row bags, reducing barrier stalls by 40\% and speeding up the kernel by 31.9\%.

Second, even when warps within a CTA have balanced workloads, the warps that reach the barrier first still wait for the remaining warps of their CTA to arrive, and during that wait the SM schedulers may lack other ready warps either because too few CTAs reside on the SM or because the other warps on the SM are also stalled (e.g., waiting on memory accesses).
We test this case on the H20 with a 1D stencil kernel: in each iteration, each warp loads a block of elements from global memory into shared memory, updates each element from its neighbor in shared memory, and synchronizes at a barrier.
Marking the instructions after the barrier up to the next iteration's shared-memory store lets the warps that reach the barrier first start loading their elements for the next iteration while waiting for the rest of their CTA.
When a single 32-warp CTA runs on each SM, no other CTAs are present to supply ready warps during the barrier wait, and marking speeds up the kernel by 7.2\%.
By contrast, when eight 8-warp CTAs share each SM, ready warps from other CTAs keep the issue slots active in 76\% of cycles, so although \EX{} still reduces barrier stalls by 31\%, kernel execution time remains unchanged.
Even when multiple CTAs share each SM, however, if their warps are stalled on memory accesses, the schedulers still lack ready warps during barrier waits: in PyTorch's \sass{F.dropout} kernel, where warps spend $50.5\%$ of their stall cycles waiting on global memory, marking with \EX{} speeds up the kernel by 6.0\%.

\section{Programmable Warp Priority (\START, \TILE, \END)}
\label{sec:priority}

\subsection{Hardware Mechanism}
\label{sec:tile-behavior}

\begin{figure}[t]
\centering
\includegraphics[width=\columnwidth]{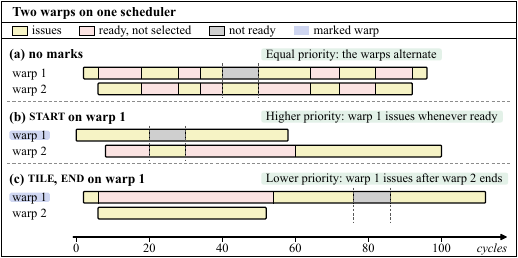}
\caption{Measured issue timeline of two warps on one scheduler of an H20: (a)~without marks, (b)~with \START{}
on warp~1, and (c)~with \TILE{} and \END{} on warp~1. In all three, warp~1 runs the same code and is not ready for
12~cycles between the dashed lines.}
\label{fig:priority}
\end{figure}

To determine how \START{}, \TILE{}, and \END{} affect warp scheduling, we construct a SASS microbenchmark in which two warps of a CTA share one warp scheduler (Fig.~\ref{fig:priority}).
Each warp executes a sequence of instructions with the yield bit set so that the scheduler can switch between ready warps after any instruction, and we record the issue cycle of each instruction using the hardware clock.
We compare running this benchmark without marks (Fig.~\ref{fig:priority}(a)), with \START{} on warp~1's first instruction (Fig.~\ref{fig:priority}(b)), and with \TILE{} and \END{} on warp~1's first two instructions (Fig.~\ref{fig:priority}(c)), while leaving warp~2 unmarked.

Without marks (Fig.~\ref{fig:priority}(a)), both warps have equal priority: the scheduler alternates between them whenever both are ready, issues from warp~2 while warp~1 is not ready (cycles~40 to~50), and finishes both warps at nearly the same time (cycles~92 and~96).
With \START{} on warp~1 (Fig.~\ref{fig:priority}(b)), warp~1 has higher priority than warp~2: the scheduler issues exclusively from warp~1 whenever it is ready (cycles~0 to~20 and~30 to~58), and switches to warp~2 only while warp~1 is not ready (cycles~20 to~30) or after warp~1 finishes (cycles~60 to~100).
When we also place \END{} in warp~1 after \START{}, the two warps resume alternating right after \END{}, showing that \END{} returns warp~1 to the same priority as warp~2.
Marking warp~1 with \TILE{} alone leaves the two warps alternating at equal priority, just as without marks.
When \TILE{} is followed by \END{} (Fig.~\ref{fig:priority}(c)), however, after warp~1 executes \TILE{} and \END{} by cycle~6, it drops to lower priority than warp~2: it waits while the unmarked warp~2 issues all of its instructions uninterrupted (cycles~6 to~52), and resumes only after warp~2 finishes (cycles~54 to~112).

These timelines show that the warp scheduler maintains three priority levels (high, normal, and low) and that \START{}, \TILE{}, and \END{} move a warp between them.
By default, all warps have normal priority.
Executing \START{} raises a warp to high priority so that it issues ahead of all normal-priority warps until it executes \END{}, which returns it to normal priority.
Conversely, after a warp executes \TILE{}, executing \END{} demotes it to low priority so that it issues behind all normal-priority warps until a subsequent \TILE{} restores it to normal priority (or \START{} raises it to high priority).
Together, these marks provide instruction-level programmable warp priority in the warp scheduler, allowing software to boost or demote a warp across a code region.
Unlike \EX{}, which depends on the instruction pipeline, \START{}, \TILE{}, and \END{} act on the warp scheduler and behave identically across all instructions and GPU architectures.
In NVIDIA's libraries (\S\ref{sec:corpus}), \sass{cuDNN} places \START{} and \END{} around batches of 4 to 16 texture loads (\sass{TLD}) to keep a warp at high priority until the entire batch issues, whereas \TILE{} is never used.

\subsection{Performance Optimization}
\label{sec:tile-perf}

\begin{figure}[t]
\centering
\includegraphics[width=\columnwidth]{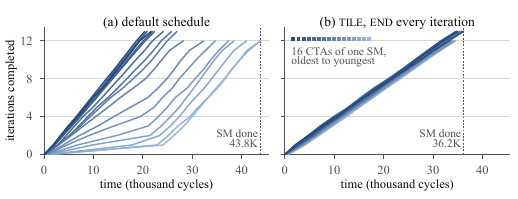}
\caption{Progress of the 16 CTAs on one H20 SM in the CUDA SDK \sass{BlackScholes} kernel.}
\label{fig:tile}
\end{figure}

Because \START{}, \TILE{}, and \END{} control warp priority in the warp scheduler, they can be used to improve performance when the hardware's default scheduling order leaves the SM underutilized.
When multiple CTAs share an SM, the default warp scheduler follows a greedy-then-oldest (GTO) policy~\cite{rogers12ccws}, selecting ready warps from older CTAs ahead of those from younger CTAs.
As a result, the older CTAs complete and exit the SM early, and when no new CTAs are scheduled onto the SM, the remaining younger CTAs execute with too few active warps to hide memory latency.
Placing \TILE{} and \END{} in each loop iteration of a kernel eliminates this bottleneck.
When warps from older CTAs execute \TILE{} and \END{}, they drop to low priority, so the warp scheduler switches to warps from younger CTAs that remain at normal priority until they also execute \TILE{} and \END{}.
In the next iteration, \TILE{} restores each warp to normal priority before \END{} demotes it again, forcing all CTAs on the SM to advance at the same pace and maintaining enough active warps throughout execution to hide memory latency.

We test this effect on the H20 with the \sass{BlackScholes} option-pricing kernel from the CUDA SDK, running 16 four-warp CTAs on each SM for 12 or 13 iterations (Fig.~\ref{fig:tile}).
In each iteration, each thread loads option parameters from global memory, evaluates the Black-Scholes formula, and writes the results back to global memory.
Without marks (Fig.~\ref{fig:tile}(a)), the older CTAs run far ahead of the younger CTAs: by the time the oldest CTA finishes all its iterations, the youngest CTA has not yet finished its first, and the SM finishes after 43.8K~cycles.
With \TILE{} and \END{} placed on the first two instructions of each iteration (Fig.~\ref{fig:tile}(b)), all 16 CTAs advance together and the SM finishes after 36.2K~cycles, 21\% faster.

\section{Post-Compilation Binary Optimizer}
\label{sec:pass}

Sections~\ref{sec:exempt} and~\ref{sec:priority} show how \EX{}, \TILE{}, and \END{} can optimize GPU kernel performance.
However, these optimizations cannot be applied to existing CUDA programs by modifying source code or the compiler, because neither CUDA nor PTX exposes SASS control codes and \sass{ptxas} is closed-source.
We therefore implement a post-compilation binary optimizer that runs after \sass{ptxas} and rewrites the compiled \sass{cubin} directly (Fig.~\ref{fig:pass}).
The optimizer sets \EX{} (Fig.~\ref{fig:pass}, \textcircled{\footnotesize 1}) and \TILE{} and \END{} (Fig.~\ref{fig:pass}, \textcircled{\footnotesize 2}) in the control codes of existing instructions without adding, removing, or reordering any instruction.
We also provide a lightweight wrapper around \sass{ptxas} that intercepts each \sass{ptxas} call during \sass{nvcc} or JIT compilation and runs the optimizer on the generated \sass{cubin}, requiring no changes to build scripts.

\subsection{\EX{} Placement}
\label{sec:pass-exempt}

\begin{figure}[t]
\centering
\includegraphics[width=\columnwidth]{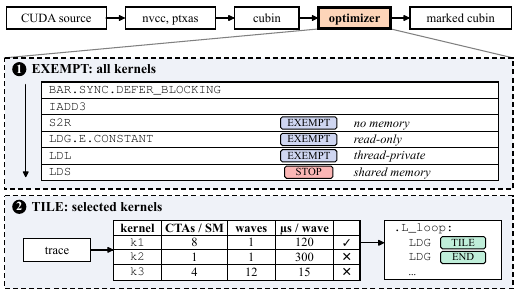}
\caption{The post-compilation binary optimizer: \EX{} in every kernel from the cubin alone, \TILE{} and \END{} in kernels selected from
one trace of the original program.}
\label{fig:pass}
\end{figure}

As shown in \S\ref{sec:exempt-opt}, after a deferred \sass{BAR}, an early warp can execute blocked instructions during the barrier wait as long as they do not conflict with other warps across the barrier.
To place \EX{}, the optimizer scans the instructions following each \sass{BAR.SYNC.DEFER\_BLOCKING} in every kernel (Fig.~\ref{fig:pass}, \textcircled{\footnotesize 1}).
Arithmetic instructions that are not blocked by a pending barrier (such as \sass{IADD3}) need no mark and are skipped.
Among blocked instructions, the optimizer marks those whose opcodes alone guarantee no conflict with other warps: instructions that do not access memory (such as \sass{S2R}, \sass{MUFU}, \sass{F2F}, and branches), read-only constant loads (\sass{LDG.CONSTANT}, \sass{LDC}), and thread-private local-memory accesses (\sass{LDL}, \sass{STL}).
If a marked instruction belongs to the \texttt{Other} category in Table~\ref{tab:block} (such as unconditional \sass{BRA} or \sass{NOP}), the optimizer sets the yield bit alongside \EX{}, and when a marked instruction is a branch, scanning continues along both branch targets (visiting each instruction at most once per barrier).

Scanning stops at the first blocked instruction outside these categories, such as the shared-memory load \sass{LDS} in Fig.~\ref{fig:pass}, a general global-memory access (\sass{LDG}, \sass{STG}), or a synchronization instruction (\sass{BAR}, \sass{MEMBAR}, \sass{DEPBAR}).
Unlike constant or local memory, a general global or shared-memory access may target an address read or written by another warp across the barrier.
Because the optimizer sees only the compiled \sass{cubin} without source-level pointer information, it cannot rule out such conflicts and leaves the instruction unmarked so that the warp waits for the barrier there.
Once the warp stalls at that unmarked instruction, marking any subsequent instruction has no effect, so the optimizer stops scanning for the current barrier.
Because these rules depend only on the SASS opcodes in the \sass{cubin}, the optimizer applies \EX{} statically to all kernels without running the program.

\subsection{\TILE{} and \END{} Placement}
\label{sec:pass-tile}

As shown in \S\ref{sec:tile-perf}, \TILE{} and \END{} improve performance by keeping multiple CTAs on an SM advancing at the same pace so that older CTAs do not exit early and leave the SM with too few active warps.
The optimizer therefore applies \TILE{} and \END{} only to kernels where multiple CTAs share each SM, each launch spans only a few waves (full sets of CTAs that reside on the GPU at once), and each wave runs long enough.
However, these three conditions depend on runtime launch configurations and input sizes, which are not available in the compiled \sass{cubin}.
To check them, the optimizer collects a single execution trace of the program using Nsight Systems (\sass{nsys}) and records each kernel's resident CTAs per SM (\texttt{CTAs\;/\;SM}), number of waves (\texttt{waves}), and duration per wave (\texttt{$\mu$s\;/\;wave}), as shown in Fig.~\ref{fig:pass} (\textcircled{\footnotesize 2}).
The optimizer then selects kernels that spend at least half of their GPU time in launches with \texttt{CTAs\;/\;SM}~${\ge}2$, \texttt{waves}~${\le}2$, and \texttt{$\mu$s\;/\;wave}~${\ge}20$.
For example, in Fig.~\ref{fig:pass} (\textcircled{\footnotesize 2}), kernel \sass{k1} meets all three thresholds and is selected, whereas \sass{k2} (1 CTA per SM) and \sass{k3} (12 waves of $15\,\mu\text{s}$) are excluded.

In each selected kernel, the optimizer finds every loop and places \TILE{} and \END{} on the first two instructions of its loop body (Fig.~\ref{fig:pass}, \textcircled{\footnotesize 2}, \sass{.L\_loop}).
If an instruction at the start of the loop body already uses bits \texttt{124:122} for \EX{} or operand-reuse flags, the optimizer skips it and marks the next available instruction in the loop body without overwriting existing flags or marks.

\section{Evaluation}
\label{sec:eval}

\subsection{Experimental Setup}
\label{sec:eval-setup}

\begin{table}[t]
\caption{Evaluated GPUs and marks placed by the optimizer.}
\label{tab:gpus}
\centering\footnotesize
\setlength{\tabcolsep}{2.8pt}
\begin{tabular}{@{}llrrrcr@{}}
\toprule
 & & & & \multicolumn{2}{c}{\EX{} marks} & \TILE{} kernels \\
\cmidrule(lr){5-6}
GPU & Arch & SMs & Mem. & llama.cpp & vLLM & (vLLM) \\
\midrule
H20 & Hopper & 78 & 96\,GB & 168\,k & 399\,k & 2 \\
H100 & Hopper & 132 & 80\,GB & 168\,k & 399\,k & 2 \\
A800 & Ampere & 108 & 80\,GB & 154\,k & 367\,k & 3 \\
RTX 4090 & Ada & 128 & 24\,GB & 140\,k & 348\,k & 2 \\
RTX PRO 6000 & Blackwell & 188 & 96\,GB & 177\,k & 400\,k & 2 \\
GB10 & Blackwell & 48 & 128\,GB & 165\,k & 398\,k & 1 \\
\bottomrule
\end{tabular}
\end{table}

We evaluate the optimizer across six NVIDIA GPUs spanning the Ampere, Ada, Hopper, and Blackwell architectures, as shown in Table~\ref{tab:gpus}.
We test two production LLM inference frameworks, \sass{llama.cpp}~\cite{llamacpp} and \sass{vLLM}~\cite{kwon23vllm}, on the dense Qwen3-8B model and the mixture-of-experts Qwen3-30B-A3B model~\cite{yang25qwen3} with 4-bit quantization, using Q4\_K\_M in \sass{llama.cpp} and AWQ~\cite{lin24awq} and GPTQ~\cite{frantar23gptq} in \sass{vLLM}.
On each GPU, we use the original build compiled by \sass{nvcc} as the baseline and compare it against the build processed by our optimizer.
In \sass{llama.cpp}, the optimizer places 140k to 177k \EX{} marks across its 144 \sass{cubin}s, as listed in Table~\ref{tab:gpus}.
Because the \TILE{} selection in \S\ref{sec:pass-tile} selects only 3 of the 71 kernels that \sass{llama.cpp} executes on the H20, which together take 1\% of the GPU time, we apply \EX{} alone to \sass{llama.cpp}.
In \sass{vLLM}, the optimizer places 348k to 400k \EX{} marks across the compiled \sass{cubin}s and applies \TILE{} and \END{} to the 20 to 22 loop bodies of the 1 to 3 kernels selected on each GPU.
Across all GPUs, the optimized builds pass all 18505 \sass{llama.cpp} unit tests and give the same result as the baseline on each of the 10267 \sass{vLLM} unit tests.

\subsection{Kernel-Level Performance}
\label{sec:eval-kernel}

We first evaluate the speedup of the individual kernels invoked during the LLM inference workloads above, as shown in Table~\ref{tab:kernels} for the H20.
In \sass{llama.cpp}, the largest gains come from the quantized matrix-multiplication kernel \sass{mul\_mat\_q}, which synchronizes warps with a barrier in each main-loop iteration before accessing shared memory.
Marking blocked instructions after each barrier lets early warps execute subsequent instructions during the barrier wait until the first shared-memory access.
As a result, its small-tile variants \sass{mul\_mat\_q<Q4\_K,32>} and \sass{mul\_mat\_q<Q6\_K,32>} speed up by $+4.7\%$ and $+3.2\%$, while its large-tile variants \sass{mul\_mat\_q<Q4\_K,128>} and \sass{mul\_mat\_q<Q6\_K,128>} speed up by $+2.1\%$ and $+2.7\%$.
The small-tile variants gain more because each iteration performs less computation per barrier, making barrier stalls a larger fraction of execution time.
In \sass{mul\_mat\_q<Q6\_K,128>}, marking local-memory spill accesses \sass{LDL} and \sass{STL} increases the marked instructions from 40 to 494 and raises its speedup from $+0.4\%$ to $+2.7\%$, as early warps would otherwise stop at the first register spill.
Beyond matrix multiplication, the normalization and positional-encoding kernels \sass{rms\_norm\_f32} and \sass{rms\_norm\_mul\_rope\_f32} speed up by $+1.4\%$ and $+1.6\%$ by executing blocked instructions such as \sass{MUFU.RSQ} early after their reduction barriers.
\begin{table}[t]
\caption{Speedup of production kernels on an H20 with \EX{} alone and with \TILE{} and \END{} on the selected kernels.}
\label{tab:kernels}
\centering\footnotesize
\setlength{\tabcolsep}{5.5pt}
\begin{tabular}{@{}llrr@{}}
\toprule
Framework & Kernel & \EX{} & +\TILE{} \\
\midrule
\sass{llama.cpp} & \sass{mul\_mat\_q<Q4\_K,32>} & $+$4.7\% & -- \\
\sass{llama.cpp} & \sass{mul\_mat\_q<Q6\_K,32>} & $+$3.2\% & -- \\
\sass{llama.cpp} & \sass{mul\_mat\_q<Q4\_K,128>} & $+$2.1\% & -- \\
\sass{llama.cpp} & \sass{mul\_mat\_q<Q6\_K,128>} & $+$2.7\% & -- \\
\sass{llama.cpp} & \sass{rms\_norm\_f32} & $+$1.4\% & -- \\
\sass{llama.cpp} & \sass{rms\_norm\_mul\_rope\_f32} & $+$1.6\% & -- \\
\midrule
\sass{vLLM} & \sass{Marlin}~\cite{frantar25marlin} GEMM ($M{=}64$) & $+$1.4\% & -- \\
\sass{vLLM} & \sass{Marlin} MoE ($M{=}32$) & $+$0.6\% & $+$4.8\% \\
\sass{vLLM} & \sass{Marlin} MoE ($M{=}128$) & $+$0.5\% & $+$5.4\% \\
\sass{vLLM} & \sass{Marlin} MoE ($M{=}256$) & $+$0.8\% & $+$0.8\% \\
\bottomrule
\end{tabular}
\end{table}

In \sass{vLLM}, Table~\ref{tab:kernels} shows the speedup of the dense \sass{Marlin} GEMM kernel and the \sass{Marlin} MoE kernel across different numbers of input tokens $M$.
Applying \EX{} alone speeds up \sass{Marlin} GEMM by $+1.4\%$ at $M{=}64$ and \sass{Marlin} MoE by $+0.6\%$ at $M{=}32$, $+0.5\%$ at $M{=}128$, and $+0.8\%$ at $M{=}256$, as early warps execute blocked address and control-flow instructions after each main-loop barrier.
When \TILE{} and \END{} are also placed on the selected \sass{Marlin} MoE kernel, which runs 3 to 4 CTAs of 128 threads per SM for $M \le 128$, keeping the CTAs on each SM advancing at the same pace as shown in \S\ref{sec:tile-perf} boosts the kernel speedup to $+4.8\%$ at $M{=}32$ and $+5.4\%$ at $M{=}128$.
By contrast, at $M{=}256$, \sass{Marlin} MoE uses a larger tile that fits only a single 256-thread CTA per SM, so \TILE{} and \END{} do not apply and the speedup remains $+0.8\%$ from \EX{} alone.

\subsection{End-to-End Inference Performance}
\label{sec:eval-e2e}

\begin{figure}[t]
\centering
\includegraphics[width=\columnwidth]{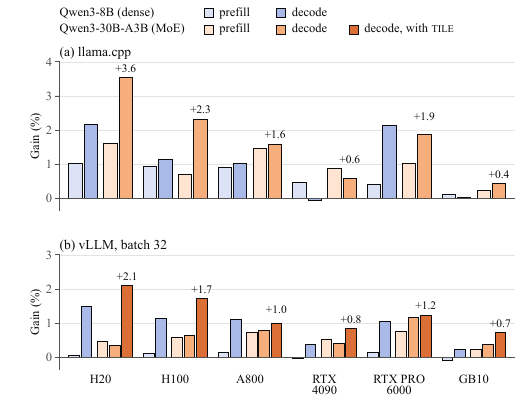}
\caption{End-to-end speedup of the optimizer over the baseline on dense Qwen3-8B and MoE Qwen3-30B-A3B for prefill and 32-sequence decoding in (a)~\sass{llama.cpp} with \EX{} and (b)~\sass{vLLM} with \EX{} alone and with \TILE{} and \END{} on the selected kernels.}
\label{fig:e2e}
\end{figure}

We next evaluate the end-to-end inference speedup of \sass{llama.cpp} and \sass{vLLM} across all six GPUs on the dense Qwen3-8B and MoE Qwen3-30B-A3B models for prefill of a 512-token prompt and decoding with 32 parallel sequences, as shown in Fig.~\ref{fig:e2e}.
In \sass{llama.cpp} (Fig.~\ref{fig:e2e}(a)), the optimizer speeds up the dense model by up to $+1.0\%$ in prefill and $+2.2\%$ in 32-sequence decoding, and speeds up the MoE model by up to $+1.6\%$ in prefill and $+3.6\%$ in 32-sequence decoding ($+3.6\%$ on the H20, $+2.3\%$ on the H100, $+1.9\%$ on the RTX PRO~6000, and $+1.6\%$ on the A800).
The speedup is largest in MoE 32-sequence decoding because, unlike prefill and dense decoding where all tokens multiply the same weight matrices using large-tile \sass{mul\_mat\_q} variants, MoE decoding routes the 32 tokens across different experts.
Each expert therefore processes only a few tokens and invokes the small-tile variants (such as \sass{mul\_mat\_q<Q4\_K,32>}, which accounts for $47.7\%$ of GPU time on the H20), where each iteration performs less computation per barrier so hiding barrier stalls with \EX{} saves a larger fraction of execution time.

In \sass{vLLM} (Fig.~\ref{fig:e2e}(b)), on the dense model, the \TILE{} selection in \S\ref{sec:pass-tile} selects no kernels because the dense \sass{Marlin} GEMM kernel runs only one CTA per SM.
Applying \EX{} alone changes prefill by less than $0.2\%$ and speeds up 32-sequence decoding by up to $+1.5\%$ ($+1.5\%$ on the H20, $+1.1\%$ on the H100 and A800, and $+1.0\%$ on the RTX PRO~6000), as decoding uses smaller matrix tiles where barrier stalls account for a larger share of execution time.
On the MoE model, applying \EX{} alone speeds up prefill by up to $+0.8\%$ and 32-sequence decoding by $+0.4\%$ to $+1.2\%$.
When \TILE{} and \END{} are also placed on the selected kernels, such as the \sass{Marlin} MoE decode instances that run 3 to 4 CTAs per SM (Table~\ref{tab:kernels}), keeping the CTAs on each SM advancing at the same pace boosts the 32-sequence decode speedup to $+2.1\%$ on the H20 (and $+2.3\%$ at batch~64), $+1.7\%$ on the H100, $+1.2\%$ on the RTX PRO~6000, and $+1.0\%$ on the A800.

\section{Related Work}
\label{sec:related}

\textbf{Reverse engineering NVIDIA GPUs.}
Microbenchmarking and binary analysis have revealed the pipelines, control codes, and encodings of NVIDIA GPUs~\cite{wong10demystify,jia18volta,jia19turing,luo24hopper,huerta25cores,hayes19decoding,documentsass,avalos27eoss,gai26f2asm} and improved GPU simulators~\cite{khairy20accelsim}, but none explains bits \texttt{124:122} (\S\ref{sec:corpus}).

\textbf{SASS-level optimization.}
Experts hand-tune SASS with third-party assemblers~\cite{gray14maxas,zhang17baremetal,yan20tensorcore}, DeepGEMM patched yield and reuse bits of compiled \sass{FFMA}s~\cite{deepgemm}, and CuAsmRL searches SASS schedules per kernel~\cite{he25cuasmrl}.
Our optimizer only sets marks, using fixed rules.

\textbf{Barrier synchronization.}
\EX{} realizes the fuzzy barrier~\cite{gupta89fuzzy} in hardware.
Because the warps of a CTA often progress unevenly~\cite{xiang14warpdiv,liu16baws}, prior work proposes hardware schedulers that accelerate critical warps or deprioritize waiting ones~\cite{lee14caws,lee15cawa,liu16baws,eltantawy18bows}, or separates barrier arrival from waiting (e.g., PTX \sass{bar.arrive} or \sass{mbarrier}~\cite{nvidia_ptx}) in warp-specialized kernels~\cite{bauer11cudadma,bauer14singe,shah24fa3,chen26tawa,soi26twill}, whose synchronization is hard to verify~\cite{sharma15weft}.
\EX{} needs neither new hardware nor kernel rewrites.

\textbf{Warp scheduling.}
Two-level~\cite{narasiman11twolevel}, cache-conscious~\cite{rogers12ccws}, CTA-aware~\cite{jog13owl}, CTA-throttling~\cite{kayiran13dyncta,lee14tbs}, and progress-aware~\cite{anantpur15pro,shen26paws} schedulers need new hardware, unlike \START{}, \TILE{}, and \END{}.

\section{Conclusion}
\label{sec:concl}

This paper presented the first systematic study of an unexplored 3-bit SASS control-code field in NVIDIA GPUs from Volta to Blackwell.
Using microbenchmarks, we showed that this field controls two independent hardware mechanisms: barrier exemption (\EX{}), which issues blocked instructions early during a barrier wait, and programmable warp priority (\START{}, \TILE{}, and \END{}), which moves warps across three priority levels in the warp scheduler.
We further implemented a post-compilation binary optimizer that rewrites compiled \sass{cubin}s directly, accelerating representative kernels by up to 31.9\%, production kernels in \sass{llama.cpp} and \sass{vLLM} by up to 5.4\%, and end-to-end LLM inference by up to 3.6\% across six GPUs.

\bibliographystyle{IEEEtran}
\bibliography{refs}
\end{document}